\documentclass[letterpaper]{IEEEtran}
\usepackage{amsmath}
\usepackage{graphicx,subfigure,comment}
\usepackage[table,dvipsnames]{xcolor}
\usepackage{url} 
\usepackage{dsfont}
\usepackage{cite}
\usepackage{tabularx}
\usepackage{dblfloatfix}

\newcommand{\Fig}[1]{Fig.~\ref{fig:#1}}

\newcommand{\Sec}[1]{Sec.~\ref{sec:#1}}

\begin{document}

\title{
Service Shifting: a Paradigm\\for Service Resilience in 5G
} 

\author{
    \IEEEauthorblockN{Francesco Malandrino\IEEEauthorrefmark{1}\IEEEauthorrefmark{2}\IEEEauthorrefmark{3}, Carla Fabiana Chiasserini\IEEEauthorrefmark{2}\IEEEauthorrefmark{1}\IEEEauthorrefmark{3}, Giada Landi\IEEEauthorrefmark{4}}
    \\
    \IEEEauthorblockA{\IEEEauthorrefmark{1}CNR-IEIIT, Italy;}
    \IEEEauthorblockA{\IEEEauthorrefmark{2}Politecnico di Torino, Italy;}
    \IEEEauthorblockA{\IEEEauthorrefmark{3}CNIT, Italy;}
    \IEEEauthorblockA{\IEEEauthorrefmark{4}NextWorks s.r.l., Italy
    \\francesco.malandrino@ieiit.cnr.it, chiasserini@polito.it, g.landi@nextworks.it}
}

\maketitle

\pagestyle{plain}

\begin{abstract}
Many real-world services can be provided through multiple virtual network function (VNF) graphs, corresponding, e.g., to high- and low-complexity variants of the service itself. Based on this observation, we extend the concept of service scaling in network orchestration to {\em service shifting}, i.e., upgrading or downgrading the VNF graph to use among those implementing the same service. Service shifting can serve multiple goals, from reducing operational costs to reacting to infrastructure problems. Furthermore, it enhances the flexibility of service-level agreements between network operators and third party content providers (``verticals''). In this paper, we introduce and describe the service shifting concept, its benefits, and the associated challenges, with special reference to how service shifting can be integrated within real-world 5G architectures and implementations. We conclude that existing network orchestration frameworks can be easily extended to support service shifting, and its adoption has the potential to make 5G network slices easier for the operators to manage under high-load conditions, while still meeting the verticals' requirements.
\end{abstract}

\section{Introduction}
\label{sec:intro}

5G networks are built {\em for services}, not merely for connectivity. Third-party providers, called {\em verticals} (e.g., automotive industries, e-health companies, and media content providers), will purchase from mobile operators the networking and processing capabilities necessary to provide their services. Such services will concurrently run on the mobile operator's infrastructure, which will support their diverse requirements under the so-called {\em network slicing} paradigm~\cite{slicing1,slicing2}.

Additionally, services that have especially tight latency requirements and/or need to process extremely large amounts of traffic can leverage the so-called multi-access edge computing (MEC) paradigm. Under the MEC paradigm, computation entities (e.g., servers) are placed at the edge of the network, thus complementing Internet-based datacenters and reducing network congestion and the associated latency. By doing so, not only does MEC improve the performance of existing services, but also it enables entirely new services, including~\cite{mec} virtual and augmented reality. On the negative side, MEC servers have a limited computational and memory capabilities, which shall be shared among all deployed services.

According to the network function virtualization (NFV) technology, services are specified by verticals~\cite{slicing1,slicing2,contreras2017network} as a set of virtual network functions (VNFs) connected to form a VNF graph, along with the needed target  Key Performance Indicators (KPIs), e.g., maximum delay or minimum reliability. Operators will host the VNFs on their own infrastructure, ensuring that they are assigned enough resources for the service to meet the target KPIs while keeping operator costs as low as possible. Such a problem is known as network orchestration~\cite{slicingalgos} or VNF placement~\cite{noi-ton19}, and has been widely researched in the literature. Popular approaches and tools include queuing theory~\cite{noi-ton19}, game theory~\cite{bagaa2018coalitional}, and graph theory~\cite{draxler2018jasper}.

It is a natural and often unspoken assumption that every vertical service is associated with {\em one} VNF graph: either the service can be provided through the specified VNFs with the target KPIs, or the service deployment fails. In some cases, resource shortages are managed by limiting the damage, e.g., getting as close as possible to the target KPIs~\cite{noi-ton19} or enforcing different priorities among services; however, it is typically assumed that VNFs composing a service requested by a vertical are not changed.

On the contrary, in many real-world cases, such as those discussed in \Sec{relevance}, the same vertical service can be provided  through a full-fledged, {\em primary} VNF graph, and also in a suboptimal yet useful fashion through a different, {\em secondary} graph. The mobile operator can thus perform two additional operations when matching the services to provide with the available resources: it can {\em shift down} a certain service, dropping its primary VNF graph and deploying the secondary one in case of resource shortage, or {\em shift up} that service performing the opposite operation.

It is important to point out how service shifting is profoundly different from the familiar experience of trying to use a service, e.g., a video call, and then, if the bandwidth is insufficient, switching to a similar one, e.g., an ordinary voice call. The fundamental difference is that shifting happens within {\em the same} service, which in turn implies that:
\begin{itemize}
    \item shifting is initiated and performed by the network, and is seamless for the user;
    \item the vertical is aware of shifting decisions, and can take care of the associated non-technical aspects (e.g., discounts at billing time) with no action on the user's part.
\end{itemize}

Service shifting is also deeply different from service scaling: service scaling aims at finding enough resources to run the current VNF graph, by means of assigning more resources to currently-active servers (scale-up) or finding new servers to use (scale-out). On the other hand, service shifting is about choosing the most appropriate VNF graph to use in order to provide a given service, also considering the resources available.

In this paper, we discuss the role the service shifting operation in 5G networks, as well as the opportunities and challenges it brings. Specifically, \Sec{relevance} discusses the relevance of shifting operations, presenting several examples of services that can benefit from them. \Sec{opportunities} deals with the role of service shifting decisions in a comprehensive network orchestration strategy, and \Sec{practice} describes how it can be implemented in practice, taking a real-world 5G architecture as a reference. Finally, \Sec{results} summarizes the results of our performance evaluation, carried out through a small, yet representative, reference scenario, and \Sec{conclusion} concludes the paper.

\section{Shifting services}
\label{sec:relevance}

Shifting mostly benefits the services that leverage the MEC paradigm, i.e., services with (i) strict latency requirements, or (ii) very significant amounts of data to process as locally as possible. Different VNF graphs represent the fact that the same goal can be pursued through different strategies, associated with different resource requirements.

\begin{figure}
\centering
\subfigure[\label{fig:vnffg-sensors}]{
\includegraphics[width=1\columnwidth]{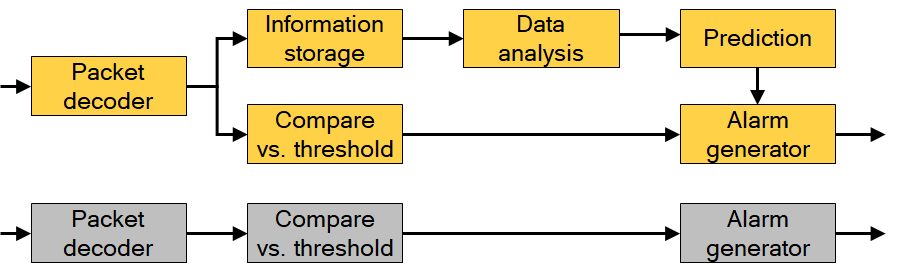}
} 
\subfigure[\label{fig:vnffg-seethrough}]{
\includegraphics[width=1\columnwidth]{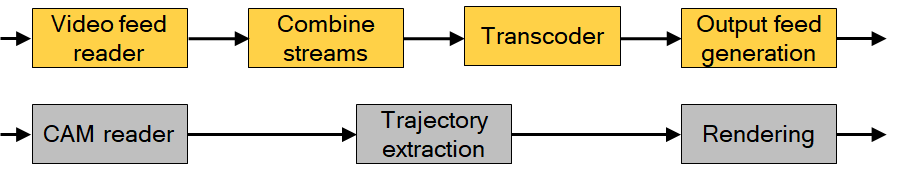}
} 
\caption{
Primary (gold background) and secondary (silver background) VNF graphs associated with a grid monitoring (a) and a bird's eye view (b) service. Note that some VNFs may be common to both graphs, as in (a).
\label{fig:vnffg}
} 
\end{figure}

A good example is the sensor monitoring service depicted in \Fig{vnffg-sensors}, presenting a power grid monitoring service~\cite[Sec.~3.4]{d11}: in ordinary conditions, sensor readings are checked against static thresholds and used for prediction. An alarm is generated if current values exceeded the static threshold, or the predicted values are detected as anomalous. However, if a resource shortage prevents the primary VNF graph from being deployed, there is a benefit in {\em at least} being able to raise an alarm if thresholds are exceeded, by implementing the bottom VNF graph in \Fig{vnffg-sensors}. Implementing such a {\em secondary} graph is preferable, for both the vertical and the mobile operator, to not implementing the service at all.

Another relevant example is the {\em Bird's eye view} service~\cite[Sec.~3.1.4]{d11}, which provides drivers (and autonomous vehicles) with a stream of real-time information about the current road conditions, including approaching vehicles/pedestrians. As depicted in \Fig{vnffg-seethrough}, such information can be obtained from cameras mounted on vehicles or along the roads (top graph). If resources are insufficient, cooperative awareness messages (CAMs) can instead be leveraged to construct a schematic view of the positions of the nearby vehicles (bottom graph).

Note that the secondary graph is either a subset of the primary one, or it includes a (smaller) number of VNFs, each of which characterized by  lower requirements. It follows that deploying the secondary graph of a certain service {\em in lieu} of the primary one will always lead to shorter delays and a reduced resource consumption, at the price of a lesser quality of experience for the user.

\section{Applications and decision-making approaches}
\label{sec:opportunities}

Here we describe two of the main applications of service shifting, namely, reacting to resource shortage situations (\Sec{sub-shortage}) and extending the expressiveness of Service Level Agreements (SLAs) (\Sec{sub-cost}). For each application, we discuss the decision-making entities that are involved and the approaches they can take.

\subsection{Reaction to resource shortage}
\label{sec:sub-shortage}

As mentioned earlier, in 5G networks operator-owned resources (e.g., servers) are used to run vertical-specified services, i.e., the VNFs composing their VNF graph. A {\em resource shortage} situation happens when the quantity of available resources drops unexpectedly, or the traffic load grows suddenly. This can be caused by several different conditions, including:
\begin{itemize}
    \item problems in the operator infrastructure, e.g., servers breaking down or data centers becoming inaccessible due to link failures;
    \item sudden increases in traffic, including mass events (``flash crowds'');
    \item emergency situations and natural disasters, whereby parts the network infrastructure can be destroyed and network demand, by both victims and responders, increases.
\end{itemize}

In resource shortage conditions, the operator is unable to meet all target KPIs for all services. The traditional approach is to {\em re-orchestrate}~\cite{shortage-wcncw} the affected services, which include (i) moving VNFs from unavailable servers to operating ones, and (ii) scaling down the resources they are assigned. This unavoidably results in KPI targets being violated, which, in turn, may jeopardize the usefulness of the service itself, e.g., lagging video for the see-through service discussed in \Sec{relevance}.

In this context, service shifting represents a very attractive alternative to scaling down. Instead of trying to implement the primary VNF graph of a service while missing the associated KPI targets, the operator can shift down that service and provide it through its secondary VNF graph. As for choosing {\em which} services to shift down, the operator can follow several approaches, including:
\begin{itemize}
    \item revenue maximization: down-shifted services bring a reduced revenue, hence, shift down the services associated with the lowest revenue loss;
\item minimization of the user QoE degradation: down-shifted services result in  a lower user satisfaction as the quality of experience users perceive may be severely impacted, hence shift down the less popular services;
    \item  minimization of the service reaction time: re-orchestration, e.g., instantiating new VNF instances and updating routing tables, takes a non-negligible time, hence, shift down the services requiring the fewest such operations.
\end{itemize}

\subsection{Extending SLAs}
\label{sec:sub-cost}

The possibility of service shifting can be leveraged during the SLA negotiation between verticals and operators. As an example, a vertical may accept that the secondary VNF graph is used for its service for a certain fraction of requests and/or in certain times of the day, in exchange of a reduced fee. Similarly, the semantics of service priorities can be extended to mandate that a service can be shifted down only if all lower-priority services (by the same vertical) have already been shifted down.

For operators, service shifting means extending the orchestration options: in addition to VNF placement and resource assignment~\cite{noi-ton19 }, operators will be able to use shifting decisions to pursue their high-level objective to meet the  SLA commitments while minimizing costs. For verticals, service shifting is an additional way to express their needs when negotiating SLAs, thus avoiding paying for unnecessary resources or features. 

On the negative side, orchestration decisions are bound to become more complex, from several viewpoints, including the identifying the decision-making entities, provide them with the information they need, and designing swift, yet effective, algorithms for them to run. All such aspects are discussed in \Sec{practice} next.

\section{Service shifting in practice}
\label{sec:practice}

We now describe how service shifting can be implemented in real-world 5G networks. Specifically, we describe which entities will be in charge of making and enacting shifting decisions (\Sec{sub-archi}), how they will interact (\Sec{sub-flow}), and the associated challenges (\Sec{sub-challenges}).

\subsection{Shifting in 5G architectures}
\label{sec:sub-archi}

Service shifting can be viewed as an extension to traditional network orchestration, which makes orchestration decisions even more complex to handle. This further strengthens the need, recently emerged in the 5G research community, to distribute the burden of network orchestration decisions across multiple decision-making entities, working at different abstraction layers.

\begin{figure*}
\centering
\includegraphics[width=.8\textwidth]{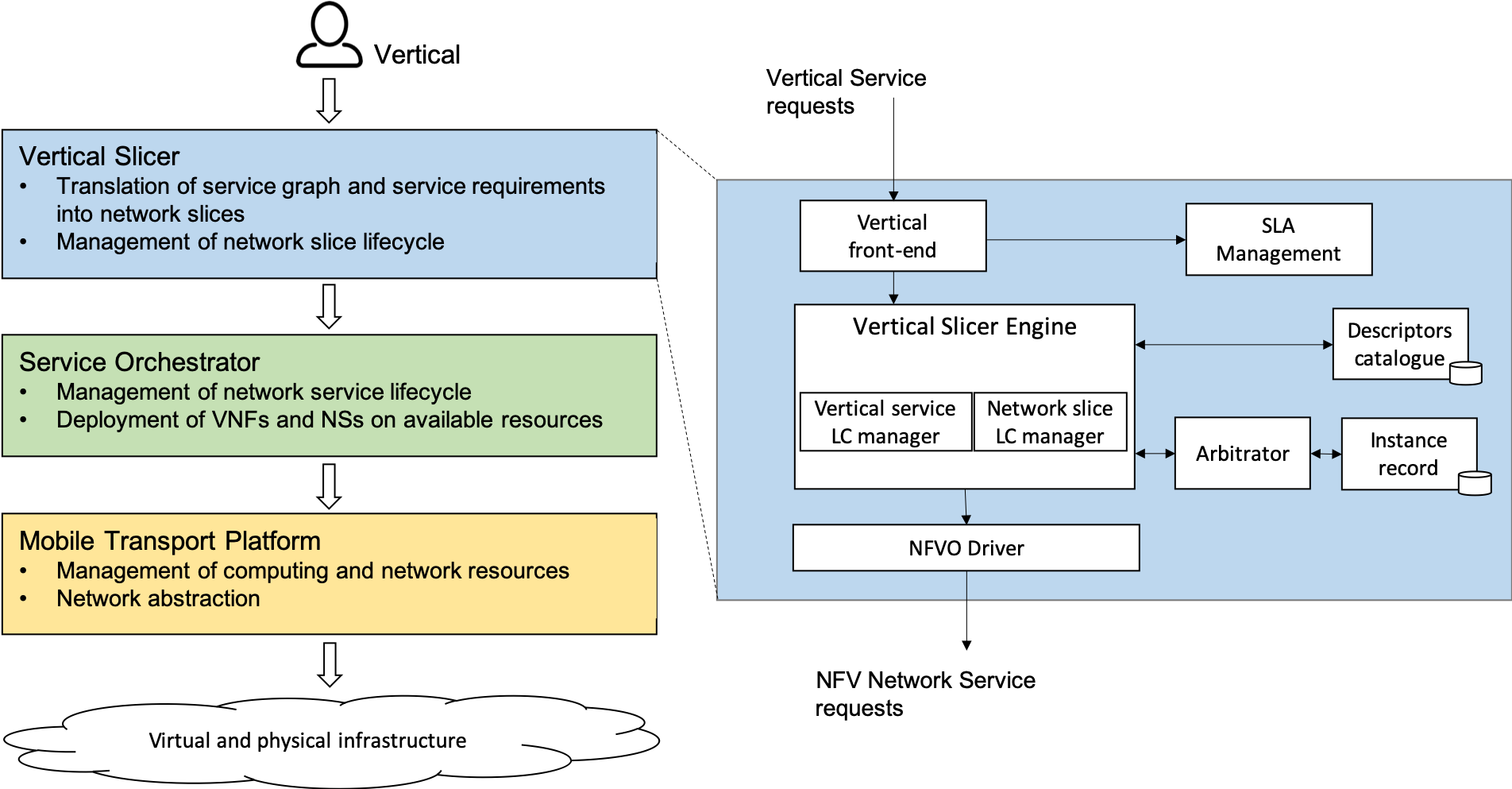}
\caption{
    The high-level architecture of the 5G-TRANSFORMER project, the interaction between decision-making entities therein, and the internal architecture of the vertical slicer.
    \label{fig:transformer}
} 
\end{figure*}

In the network management and orchestration (MANO) framework, standardized by ETSI in standard GS NFV MANO 001, virtually all network orchestration decisions are made by the NFV Orchestrator (NFVO). The NFVO takes as an input the service graphs and KPIs specified by verticals through the Operation and Business Support Services (OSS/BSS). 
Its output is represented by VNF instantiation and placement decisions, which are subsequently enacted by lower-level entities like the VNF manager (VNFM).

Several 5G-related research efforts envision alternative solutions, advocating to split the tasks assigned to the NFVO in the MANO framework between two entities: a higher-level one, making decisions on a per-service basis, and a lower-level one, working with individual VNFs with decisions more oriented to resource-based criteria. Taking the architecture proposed by the H2020 project 5G-TRANSFORMER in~\cite{5gt}, and represented in \Fig{transformer}, we can identify:
\begin{itemize}
    \item the vertical slicer (VS), translating the verticals' requirements into service graphs, also accounting for the service-level agreements (SLAs) in place;
    \item the service orchestrator (SO), taking the service graph as an input and using the network, computing and storage resources available in the infrastructure to build the network slice that will run the service.
\end{itemize}
In such a context, service shifting decisions can be made by higher-level, service-aware entities such as the VS. This avoids further increasing the burden on lower-level entities like the SO, which are already in charge of VNF placement and resource assignment.

\subsection{Making and implementing the decisions}
\label{sec:sub-flow}

As discussed earlier, in the 5G-TRANSFORMER architecture the VS will be in charge of shifting decisions. In the following, we discuss its internal architecture, also summarized in \Fig{transformer}, and how it will interact with other 5G-TRANSFORMER entities in order to make and implement the shifting decisions.

The internal architecture of the VS, summarized in the right part of \Fig{transformer}, includes multiple sub-entities, including:
\begin{itemize}
    \item the {\em arbitrator}, in charge of actually making the decisions;
    \item a SLA manager, storing information on SLA resources and tracking how they are used;
    \item catalogs and record managers, storing the information needed by the arbitrator;
    \item an engine, in charge of coordinating the work of all other VS sub-entities, as well as the life cycle (LC) of network slices;
    \item font-ends and drivers, implementing the interfaces between the VS and other 5G-TRANSFORMER entities, as well as with verticals.
\end{itemize}

Within such an architecture, adding service shifting capabilities to the VS would require four main actions. First, the vertical front-end shall be extended, in order to allow verticals to indicate multiple requirements (hence, multiple VNF graphs) for the same service. Furthermore, in order to store such VNF graphs, a new catalog shall be added, called {\em VNF graph catalog}. Additionally, the actual shifting algorithms must be implemented at the arbitrator. Finally, the NFVO driver shall be updated to convey shifting decisions from the VS to the SO.

\begin{figure}
\centering
\includegraphics[width=1\columnwidth]{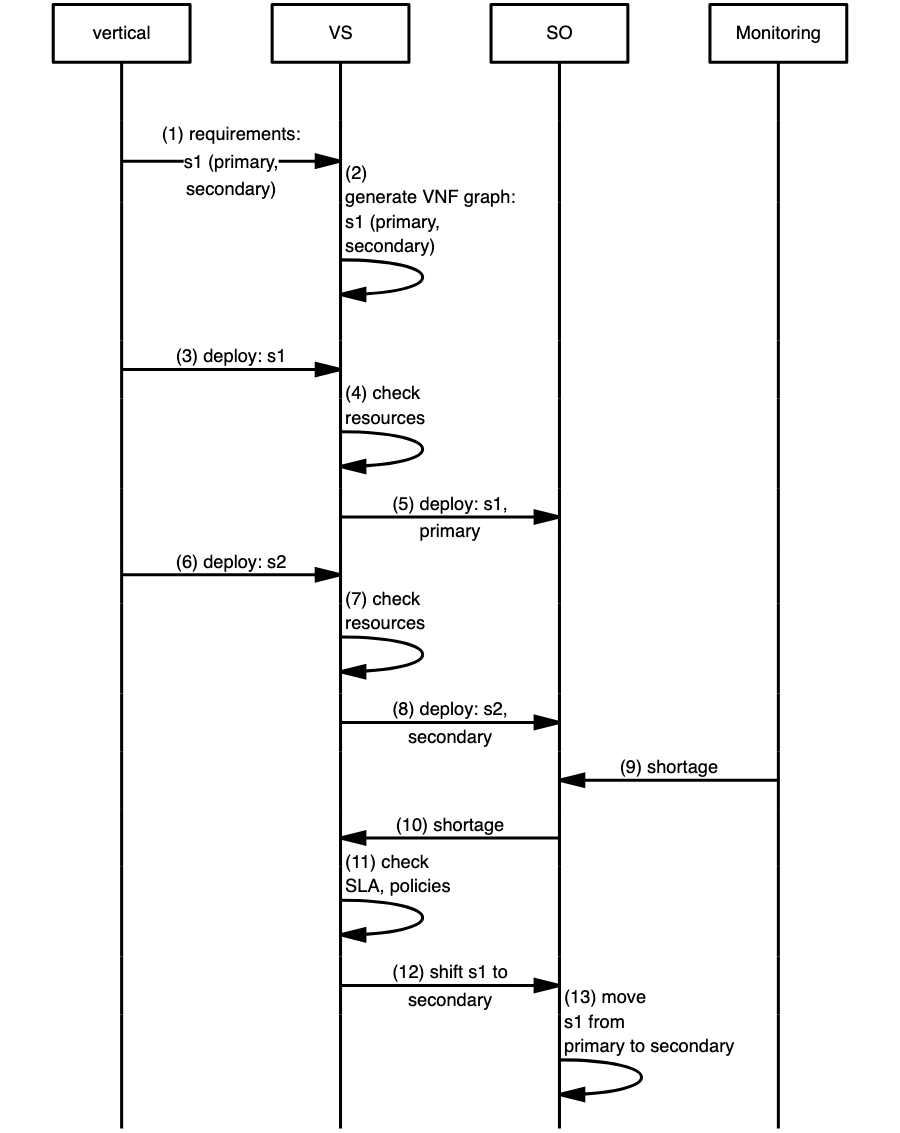}
\caption{
Making and enacting shifting decisions: interaction between  entities of the 5G network architecture.
    \label{fig:flow}
}
\end{figure}

\Fig{flow} presents a simplified vision of how 5G-TRANSFORMER entities interact when making and enacting service shifting decisions. In steps 1--2, the vertical informs the VS of the requirements associated with the different versions of its services. In steps 3--8, the vertical requests to the VS the deployment of services~$s_1$ and~$s_2$. After checking the available SLA resources, the VS decides to deploy the primary graph of~$s_1$ and the secondary  one of~$s_2$, and instructs the SO accordingly.

In step~9, the monitoring platform detects a resource shortage situation and informs the SO, which relays the warning to the VS. Such a situation requires to shift down a service, and the VS decides to shift~$s_1$ from primary to secondary. The decision is then notified to the SO, which enacts it by removing the VNFs associated with the primary graph of~$s_1$ and deploying those of the secondary graph.

\subsection{Challenges}
\label{sec:sub-challenges}

There are several challenges to tackle order to make effective service shifting decisions. Among the most significant, we discuss gathering and collecting input information, timing the decisions, and managing the transition between VNF graphs.

{\bf Input information and monitoring.}
As recalled in \Sec{sub-archi}, the VS and SO decision entities run algorithms that need to receive as input different kinds of monitoring data, related to a variety of physical and virtual components and resources, from physical infrastructures to virtual resources, up to application and service level data. The monitoring platform should be flexible enough to support different types of customizable data sources in a distributed environment. They should also  implement preliminary data elaboration tasks to efficiently deliver aggregated monitoring parameters and produce automated notifications, based on simple thresholds or more complex strategies for anomaly detection.

\begin{figure*}
\centering
\subfigure[\label{fig:compare}]{
\includegraphics[width=.3\textwidth]{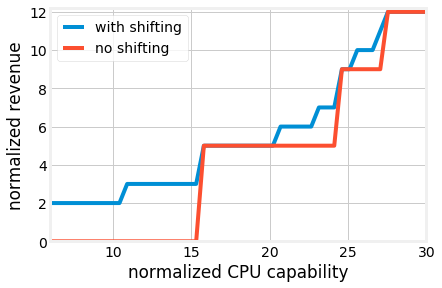}    
} 
\hspace{1cm}
\subfigure[\label{fig:bd-shifting}]{
\includegraphics[width=.3\textwidth]{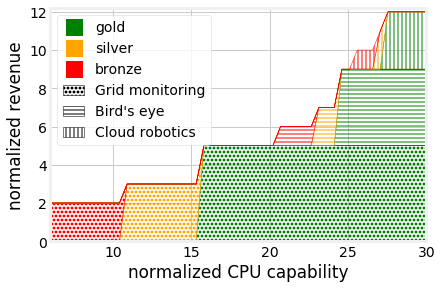}    
} 
\caption{
Small-scale example scenario: revenue obtained with and without service shifting (a) and revenue breakdown (b).
    \label{fig:performance}
}
\end{figure*}

The complexity of aggregation and elaboration of the raw monitoring data, as collected by the elementary monitoring sources, is centralized at the monitoring platform. Such processing is driven by the rules that are dynamically configured according to the network service specification, in order to detect the particular conditions triggering scaling or shifting actions. Whenever a target pattern is detected in the aggregated monitoring data, automated alerts are notified to the monitoring consumers (VS or SO) that have an active subscription for the given pattern. Notifications may be managed either through explicit messages addressed to the target entities or through a message bus approach. Starting from the received alerts, the VS or the SO will make a decision about the need of a service shifting and will trigger the required actions.

{\bf Decision timing.}
Indeed, shifting decisions are often made in resource shortage conditions, where KPI targets are being or may be violated. Therefore, service (re)deployment decisions must be made {\em and enacted} swiftly. The first requirement, i.e., that decisions be made quickly, is at odds with the complexity of the decisions to make, which include placing multiple VNFs throughout the network infrastructure. The second requirement, i.e., that decisions be enacted swiftly, is often overlooked but very important: indeed, real-world 5G deployments show VNF instantiation times of several tens of seconds~\cite{d52}.

Moreover, a full operation service also needs applications completely up and running in the new VNFs; this requires additional time due to the starting procedures of the processes and the initial configuration of the applications running in Virtual Machines (VMs) or Containers. Live migration of, e.g., VMs also brings a certain degree of delay, which may impact  the services that do not need to be shifted, but just moved to different servers. A report about live migration in OpenStack Ocata\footnote{http://superuser.openstack.org/wp-content/uploads/2017/06/ha-livemigrate-whitepaper.pdf} shows average measurements from nearly 50 seconds up to 270 seconds for the time required to migrate ``heavy''  VMs, depending on the VMs' storage strategy (i.e., local vs. shared storage) and tunneling activation. Such delays can result in non-negligible service outage times, and substantial penalties for the mobile operator.

Intuitively, taking action as early as possible is a promising way out of such a conundrum. However, early actions may turn out to be unnecessary (e.g., the traffic of a certain service did not grow as much as anticipated), or even wrong. To minimize such mishaps, several {\em traffic prediction}~\cite{scianca} techniques have been developed, typically leveraging machine learning techniques to accurately detect relevant trends.

{\bf Managing the transition between graphs.}
Shifting decisions, e.g., moving from the primary VNF graph of a service to the secondary one, require several operations on individual VNFs, e.g., deactivating those of the primary graph and activating the additional ones (if any) needed by the secondary one. The order in which such operations are performed has a significant impact on the effect of the shifting decision, and must therefore be taken into account.

One possible approach is make-before-break, i.e., first all VNFs of the secondary graph are deployed, and then those of the primary ones are removed. The main advantage of this approach is service continuity, i.e., there is no point in time at which the service is not provided. On the negative side, make-before-break means that, for a short time, both the VNF graphs will be active, and so even a shifting-down action ends up temporarily consuming more resources. This is acceptable if the action is taken early enough (e.g., thanks to effective forecast), but often infeasible in resource scarcity conditions.

The alternative approach is break-before-make, i.e., first remove the VNFs of the primary graph (that are not used by the secondary one), and then deploy those of the secondary graph (that were not already used by the primary one). This approach requires the smallest possible amount of resources, but it implies the possibility that, albeit for a limited amount of time, the service will be interrupted.
Intermediate approaches, whereby deactivation and deployment operations are interleaved, are also possible: in the 5G-TRANSFORMER architecture, it is the SO's task to decide the exact sequence of operation to perform in order to implement the shifting decisions made by the VS.

\section{Performance evaluation}
\label{sec:results}

We now quantify the benefits of service shifting by implementing the following algorithm at the VS, based on~\cite{pimrc-wp3}, operating as follows:
\begin{enumerate}
    \item the VS sorts the services in decreasing priority order;
    \item for every service:
    \begin{enumerate}
        \item start from the primary VNF graph;
        \item instruct the SO to deploy such a graph;
        \item if resources are insufficient, move to the next graph.
    \end{enumerate}
\end{enumerate}

We consider a simple, yet representative, scenario, including the two services in \Fig{vnffg} and the cloud robotics service described in~\cite[Sec.~2.4.1]{d11}. Grid monitoring has the highest priority, followed by bird's eye, and then by cloud robotics. As highlighted in~\cite{d11}, all services belong to the mission critical cluster and all require low latency and high reliability. Denoting, for simplicity, the different graphs associated to services as good, silver, and bronze, we set their the normalized requirements to (respectively) to 20, 10, and 5. Furthermore, the normalized revenues associated to the gold, silver, and bronze graphs are: $[5, 3, 2]$ for grid monitoring, $[4, 2, 1]$ for bird's eye, and $[3, 2, 1]$ for industrial robotics. 
The physical infrastructure is composed of six hosts (servers) connected in a two-layer topology reflecting the core network organization used in~\cite{topo}. For simplicity, we focus on computational capabilities alone and vary the normalized CPU available at each host between~1 and~5.

What we seek to assess is how much network shifting, i.e., the possibility to deploy silver or bronze graphs {\em in lieu} of gold ones, improves the revenue. \Fig{compare} provides a quite clear answer to our question: revenue increases as the available CPU grows, and service shifting is always associated with a higher revenue. It is even more interesting to observe, in \Fig{bd-shifting}, the services and graphs generating such a revenue, represented by patterns and colors respectively.

In the baseline case, all revenue comes from gold graphs (green areas in the plots): when the network capacity is very low, it is impossible to deploy anything; as it, the VS deploys first the gold graph of the grid monitoring service, then the gold graph of the bird's eye service, and so on. If, on the other hand, service shifting is possible, we can see that the VS is able to deploy bronze and silver graphs of different services (yellow and read areas in the plot), even when there are not enough resources for the corresponding gold graph, thereby guaranteeing a higher revenue.

\section{Conclusion}
\label{sec:conclusion}

We showed how service shifting can be beneficial in resource shortage situations, which may arise as a consequence of 5G network infrastructure issues, sudden increases in traffic demand, or emergency situations. We also identified the main challenges associated with service shifting; then, taking real-world 5G implementations as a reference, we highlighted how such challenges can be tackled without major changes to their architecture, thus making it easy to reap the benefits of service shifting. As confirmed by our performance evaluation, service shifting yields a threefold benefit: vertical requirements are satisfied in a wider range of cases, network infrastructure is better utilized, and mobile operators are able to obtain a higher revenue.

\bibliographystyle{IEEEtran}
\bibliography{refs}%

\end{document}